\magnification=\magstephalf
\baselineskip=12pt
\vsize=22.0truecm
\hsize=15.5truecm
\nopagenumbers
\parskip=0.2truecm
\def\vs{\vskip 0.2in}
\def\ts{\vskip 0.05in}

\def\n{\noindent}
\def\s{$\,$}
\font\bigbold=cmbx12

\centerline{{\bigbold Data Processing for LISA's}}

\centerline{{\bigbold Laser Interferometer Tracking System (LITS)}}
\vs
\centerline {Ronald W. Hellings}
\centerline {Jet Propulsion Laboratory, California Institute of Technology}
\centerline {Pasadena, California 91109 }

\vs

\n {\bf I.  Introduction}
\ts
The purpose of this paper is twofold.  First, we will present recent results [1] on the data processing for LISA, including algorithms for elimination of clock 
jitter noise and discussion of the generation of the data averages that will 
eventually need to be telemetered to the ground.  Second, we will argue, based 
partly on these results, that a laser interferometer tracking system (LITS) that 
employs independent lasers in each spacecraft is preferable for reasons of 
simplicity to that in which the lasers in two of the spacecraft are locked to 
the incoming beam from the third.

Before we begin, let us insert a word of introduction on this second point.  
It may seem intuitive to those who have previously worked with laboratory 
interferometers that the simplest data acquisition scheme for interferometers 
will consist in sending a signal from a central master laser out along two arms 
to end-masses where the beams are reflected with no change of phase, and the 
returning beams are recombined to form a simple Michelson interferometer.  The 
signal so formed may then be telemetered from the central spacecraft to the 
ground.  However, for LISA, there are important scientific reasons to want to 
simultaneously observe signals from a second (and probably a third) 
interferometer, with the central spacecraft of each new interferometer being one 
of the end masses of the original interferometer.  But the Michelson scheme 
won't work in these cases since the master laser is now at one of the end masses 
rather than at the central mass.  Interferometer-like signals can be formed, but 
they must each involve multiple signals, including one-way signals, from all 
three spacecraft.  A detailed discussion of this issue is given in Reference [1].

\vs
\goodbreak
\n {\bf II. Clock Jitter Cancellation}
\ts
\nobreak
Let us use notation in which the one-way signal sent from spacecraft (s/c) 2 to s/c 1 is $s_{21}(t)$.  The phase of this signal will be measured with reference to an on-board frequency standard, or `clock'.  If the laser phase noise in s/c $i$ is $p_i(t)$ and the clock jitter is $q_i(t)$, then the noise contributed to $s_{21}(t)$ by these two sources is
$$ s_{21}(t)=p_2(t-L_{12})-p_1(t)-a_{21}q_1(t) \eqno (1) $$
\n where the one-way light time in the arm connecting s/c 1 and s/c 2 is $L_{12}$ and the ratio between the clock frequency $f_1$ on s/c 1 and the laser beat frequency that is being counted on s/c 1 is $a_{21}=(\nu_2-\nu_1)/f_1$.  The term $a_{21}q_1(t)$ represents the effect of the phase jitter of the clock on the laser phase readout.

The laser-phase-noise-free interferometer-type signal originally discovered by Tinto and Armstrong [2] is 
$$ \eqalign{
X(t)=s_{12}&(t-L_{12}-2L_{13})-s_{13}(t-L_{13}-2L_{12})+s_{21}(t-2L_{13}\cr
&-s_{31}(t-2L_{12})+s_{13}(t-L_{13})-s_{12}(t-L_{12})+s_{31}(t)-s_{21}(t)}
\eqno (2) $$
\n When Eq. (1) is used in Eq. (2), the laser phase noises, $p_i(t)$, will cancel out, but a combination of the $q_i(t)$ clock jitter terms will remain and will dominate other noise sources in the detector by several orders of magnitude.

The solution to this problem employs the so-called `two-color laser' approach.  In this method, a second laser signal accompanies each main laser signal, with a frequency offset equal to the clock frequency on each spacecraft.  Thus, in addition to the signal in Eq. (1), there will be a second signal read out by each LITS,
$$ s'_{21}(t)=p_2(t-L_{12})-p_1(t)+q_2(t-L_{12})-(a_{21}+1)q_1(t) \eqno (3) $$
where the $q_2(t-L_{12})$ and $1q_1(t)$ terms arise as noise in the second laser frequencies from each spacecraft, and the $a_{21}q_1(t)$ term comes from the phase readout using the local clock.  When Eq. (3) is subtracted from Eq. (1) the resulting signal
$$ r_{21}(t)=s_{21}(t)-s'_{21}(t)=q_2(t)-q_1(t-L_{12}) \eqno (4) $$
will contain nothing but the differences of the two clock noises from the two spacecraft.  A little algebra then shows that the combination
$$ \eqalign{
\chi(t)=&X(t)-a_{12}r_{12}(t-2L_{13})+a_{13}r_{13}(t-2L_{12})
-(a_{12}+a_{21})r_{13}(t-L_{12})\cr
&+(a_{13}+a_{31})r_{12}(t-L_{12})+(a_{12}+a_{13}+a_{31})r_{32}(t)
-(a_{21}+a_{23}+a_{32})r_{12}(t)}
\eqno (5) $$
will exactly cancel the $q_i$ terms that remain in Eq. (2).  The algorithm may be extended to the $Y(t)$ and $Z(t)$ variables of Estabrook, Armstrong, and Tinto [3] by permutation of the indices $1\rightarrow2\rightarrow3$, giving clock-jitter-cleaned signals $\psi(t)$, and $\zeta(t)$.

\vs
\goodbreak
\n {\bf III. Data Requirements}
\ts
\nobreak
The cancellation of phase and clock noise in Eqs. (2) and (5) requires that the combinations of signals given on the right-hand sides of these two equations be taken instantaneously, or at least with a high enough ($\sim1\,\mu$s) time resolution that the cancellation errors are small.  However, it is not feasible to telemeter data from each spacecraft at this high rate (and such telemetry is neccessary since data from multiple spacecraft are needed in order to generate the three $\chi(t)$, $\psi(t)$, and $\zeta(t)$ signals).  A solution to this difficulty can be found by taking the averages of these equations over some sample time (say 1\s s).  The averages of the $s_{ij}$ and $r_{ij}$ signals must therefore be accumulated over the sample time, beginning at the time offset corresponding to the light-times in the arguments of the terms on the right-hand sides.  When a census is taken of all signals needed on board each spacecraft to form all three of the $\chi(t)$, $\psi(t)$, and $\zeta(t)$ combinations, a table of signal processing requirements may be formed (see Table 1).

Careful examination of the formulas for $X(t)$, $Y(t)$, and $Z(t)$ and for their clock-jitter-cleaned counterparts $\chi(t)$, $\psi(t)$, and $\zeta(t)$ concludes that the information available on board each spacecraft, as given in Table 1, may be combined into three pieces of data at each sample time that need to be communicated to the ground (see Reference [1] for details).  A reasonable number of bits to give the accumulated phase to ps resolution is 64 bits/sample.  Therefore, at a sample rate of 1 sample/s, the data requirement would be $9\times64=576\,$b/s.  It is also demonstrated in Reference [1] that locking the two end lasers to the master laser does not affect this data rate at all.  Only if the clocks on board the two end spacecraft are also and separately locked to the incoming clock signals -- {\it i.e.}, to the difference $r_{ij}$ between $s_{ij}$ and $s'_{ij}$ -- will there be a reduction of data rate to $7\times64=448\,$b/s.

\vs
\line{\hfill
\vtop{\tabskip=0pt \offinterlineskip
\halign{
\vrule width.8pt#\tabskip=1em plus 2em
&\hfil # \hfil   &\vrule width.8pt#
&\hfil # \hfil   &\vrule width.8pt#
&\hfil # \hfil   &\vrule width.8pt#
&\hfil # \hfil   &\vrule width.8pt#
&\hfil # \hfil   &\vrule width.8pt#     
\tabskip=0pt\cr
\noalign{\hrule height.8pt}
height5pt&\omit&&\omit&\omit&\omit&\omit&\omit&\omit&\omit&\cr
& signals && \omit & \omit & times & \omit & \omit &\omit &&\cr
height5pt&\omit&&\omit&\omit&\omit&\omit&\omit&\omit&\omit&\cr
\noalign{\hrule height.8pt}
height5pt&\omit&&\omit&&\omit&&\omit&&\omit&\cr
& $s_{21}$ && $t-L_{12}-2L_{23}$ && $t-L_{12}$ && $t-2L_{13}$ && $t$ &\cr
height5pt&\omit&&\omit&&\omit&&\omit&&\omit&\cr
& $r_{21}$ &&                    && $t-L_{12}$ && $t-2L_{13}$ && $t$ &\cr
height5pt&\omit&&\omit&&\omit&&\omit&&\omit&\cr
& $s_{31}$ && $t-L_{13}-2L_{23}$ && $t-L_{13}$ && $t-2L_{12}$ && $t$ &\cr
height5pt&\omit&&\omit&&\omit&&\omit&&\omit&\cr
& $r_{31}$ &&                    && $t-L_{13}$ && $t-2L_{12}$ && $t$ &\cr
height5pt&\omit&&\omit&&\omit&&\omit&&\omit&\cr
\noalign{\hrule height.8pt}
height5pt&\omit&&\omit&&\omit&&\omit&&\omit&\cr
& $s_{12}$ && $t-L_{12}-2L_{13}$ && $t-L_{12}$ && $t-2L_{23}$ && $t$ &\cr
height5pt&\omit&&\omit&&\omit&&\omit&&\omit&\cr
& $r_{12}$ &&                    && $t-L_{12}$ && $t-2L_{23}$ && $t$ &\cr
height5pt&\omit&&\omit&&\omit&&\omit&&\omit&\cr
& $s_{32}$ && $t-L_{23}-2L_{13}$ && $t-L_{23}$ && $t-2L_{12}$ && $t$ &\cr
height5pt&\omit&&\omit&&\omit&&\omit&&\omit&\cr
& $r_{32}$ &&                    && $t-L_{23}$ && $t-2L_{12}$ && $t$ &\cr
height5pt&\omit&&\omit&&\omit&&\omit&&\omit&\cr
\noalign{\hrule height.8pt}
height5pt&\omit&&\omit&&\omit&&\omit&&\omit&\cr
& $s_{13}$ && $t-L_{13}-2L_{12}$ && $t-L_{13}$ && $t-2L_{23}$ && $t$ &\cr
height5pt&\omit&&\omit&&\omit&&\omit&&\omit&\cr
& $r_{13}$ &&                    && $t-L_{13}$ && $t-2L_{23}$ && $t$ &\cr
height5pt&\omit&&\omit&&\omit&&\omit&&\omit&\cr
& $s_{23}$ && $t-L_{23}-2L_{12}$ && $t-L_{23}$ && $t-2L_{13}$ && $t$ &\cr
height5pt&\omit&&\omit&&\omit&&\omit&&\omit&\cr
& $r_{23}$ &&                    && $t-L_{23}$ && $t-2L_{13}$ && $t$ &\cr
height5pt&\omit&&\omit&&\omit&&\omit&&\omit&\cr
\noalign{\hrule height.8pt}
}}\hfill}
\n Table 1.  Signal requirements for the set $\chi(t)$, $\psi(t)$, and $\zeta(t)$.  The times listed for each signal are those at which the signal must be accumulated on board each spacecraft.

\vs
\vs
\goodbreak
\n {\bf IV. The Case for Independent Lasers}
\ts
\nobreak
In this section we will discuss the pros and cons of: 1) phase-locking the end lasers to the incoming signals, versus 2) having independent lasers in each spacecraft, each locked to its own Fabry-Perot cavity.  We make two assumptions preparatory to this discussion.  First, we assume that it is important to generate all three signals, $\chi(t)$, $\psi(t)$, and $\zeta(t)$.  The reasons for this assumption are both scientific and technical.  The scientific reasons include the fact that determining the gravitational wave polarization of short bursts requires two independent detectors and that combined polarization and directional information for signals from coalescing massive black holes is severely compromised with only a single detector present [4].  The technical reasons are derived from the fact that, in the long-wavelength limit, the response of the sum $\chi(t)+\psi(t)+\zeta(t)$ to a gravitational wave will be near zero, leaving pure instrumental noise whose level may thus be calibrated [5].  Second, we assume that it is important that any one of the three spacecraft should be able to function as the central spacecraft, providing (in the case of phase-locked end lasers) an independent master laser.  The reason for this assumption is practical; there are several failure modes in the LISA mission that would eliminate a single arm of the interferometer only.  Whichever arm is lost, if the spacecraft opposite to it can function as the vertex of the remaining two-arm interferometer, a single detector is still enabled.  With these two assumptions, we turn to the discussion of several issues related to the difference between the two LITS schemes.

\ts
\n {\it 1.  Data Rates}.  As stated in Section III, there is no data rate difference at all between the two schemes unless both the lasers and the clocks on the two slave laser spacecraft are also locked.  Even then, the reduction is only from 576\s b/s to 488\s b/s.  Locking the phases of the on-board clocks means that $r_{ij}$ will have to be monitored on each end spacecraft by subtracting the $s_{ij}$ and the $s'_{ij}$ signals detected in the optical readouts.  The $r_{ij}$ will then have to be compared with the phase of the on-board oscillator and the difference driven to zero in a feedback loop by external control of the oscillator.  Since most ultra-stable oscillators (USOs) are meant to be independent standards, this will mean designing a custom USO and employing a technique that is not a common one.

\ts
\n {\it 2.  Multiple Control Systems}.  In the phase-locked scheme, each spacecraft must be able to function, as needed, as both a master spacecraft and as a slave spacecraft.  Therefore, the control system on each spacecraft must have two very different inputs (different frequencies, different SNRs, {\it etc.}) and must be able to operate in two different modes (master and slave).  The independent laser scheme, on the other hand, requires only the usual Pound-Drever locking system which has been widely used in laboratory settings.  As discussed above, a separate clock-locking control system will be required in the locked-laser scheme if there is to be any data rate advantage in this scheme at all.  Since each spacecraft must be able to function both as master and slave, this means a two-input, two-mode control system for the local clock as well.  Finally, there is the question of initial acquisition of the signals.  Since the beam from the laser optics will be narrower than the initial pointing accuracy derived from star trackers, the acquisition will have to proceed either by expanding the initial beam or by searching through an error box.  When the end lasers are required to be phase-locked to the incoming beam, another degree of freedom in the initial signal acquisition is added.  The only reasonable way to accomplish the initial acquisition is to separate the process into two steps, but this will mean a separate frequency-independent position acquisition system and subsequent frequency acquisition system, most likely with a handoff to a more precise combined position/phase control system after signal acquisition is complete.  The independent laser scheme, by contrast, does not have to search in frequency to acquire the far spacecraft, and a single sensor could in principle provide all the pointing information required.

\ts
\n {\it 3.  Interfaces}.  Perhaps the strongest argument for independent lasers comes from the simplicity of the interfaces in this scheme versus the complexity in the case of phase-locked end lasers.  In the independent laser scheme, the laser transmitter has one self-contained function.  When it is turned on, it phase-locks to its own Fabry-Perot cavity and broadcasts a beam.  This process is completely testable in the laboratory and completely independent of any other spacecraft or any other subsystem.  It is also immediate and local rather than being delayed by a 15\s s light-time.  The laser receiver also operates independently and passively, with no contribution to any hardware control loop.  As soon as light from the far spacecraft shines on its photodiode, it finds the frequency of the strongest RF beat signal in its photodiode and measures its phase in software.  Similarly, each spacecraft in the independent laser scheme has its own clock, independently activated and internally stabilized.  In the phase-locked case, although the design of the subsystems is not yet complete, it is clear that there willl be many interdependencies.  The end spacecraft receivers must control their transmitters and return a signal before the central spacecraft's receivers see any signal at all.  There will be 15\s s time delays between spacecraft.  If there is a drop-out in the system, the entire system will have to be reinitialized and restarted.  This interdependence makes the system complicated to design and difficult to test, both elements invariably increasing cost and risk.

\ts
\goodbreak
\n {\it 4.  Frequencies}.  The one element of comparison between phase-locked and independent laser options that favors the phase-locked scheme is that the independent laser scheme will almost certainly require measurement of an order of magnitude higher fundamental beat frequency between incoming and outgoing lasers than is expected in the phase-locked case.  At the present time this means that the laser receivers in the independent laser case will need a local oscillator (LO), phase-locked to the local on-board clock, to produce a frequency $a_{21}f_1$ that beats the incoming signals down to a baseband frequency low enough that the signals can be sampled and phase-tracked without any more clock noise being added.  However, as may be seen in Eqs. (3) and (4),  the clock jitter cancellation scheme we have derived depends on the same frequency being mixed with both $s_{ij}$ and $s'_{ij}$, meaning that an LO is required anyway in both the independent laser and phase-locked laser schemes (see Reference [1] for details).  So the only real difference between the receivers in the two schemes is the frequency of the LO, which would be $\sim300\ $MHz for the independent laser case and $\sim10\ $MHz for the phase-locked laser case.  The higher frequency is not a real problem and a breadboard version of a low-noise 300\s MHz LO that is phase-locked to an external frequency standard has been designed and built [6].

\vs
\goodbreak
\n {\bf V. Conclusions}
\ts
\nobreak
In summary, the clock jitter noise, which would otherwise raise the LISA noise by three or four orders of magnitude, can be eliminated by use of time-domain algorithms that are free of the singlarities that arose in the previous frequency-domain method [7].  This cancellation, like the frequency-domain method that preceeded it, requires a second laser frequency in the main beams of the interferometer and requires additional data processing on-board each spacecraft, but the data rate to the ground is unaffected.  The data rate we find, however, is somewhat higher than has previously been noted since we have, for the first time, worked out explicitly how to generate the phase-noise-free signals by collecting data from the appropriate spacecraft.  Finally, we have analysed how the two schemes, phase-locked end-lasers and all-independent lasers, compare, both with regard to data rate and simplicity of operation, and have come to the conclusion that the small data-rate reduction enabled by the phase-locked end-laser scheme does not seem to justify the complications and risk that this method entails.

\vs

\n {\bf References:}
\ts

\font\bo=cmbx10

\n
\hangindent=0.2in
\hangafter=1
[1] R.W. Hellings, Phys.\ Rev. {\bo D}, submitted (2000).

\n
\hangindent=0.2in
\hangafter=1
[2] M. Tinto \& J.W. Armstrong, Phys.\ Rev. {\bo D59}, 102003 (1999).

\n
\hangindent=0.2in
\hangafter=1
[3] F.B. Estabrook, M. Tinto, \& J.W. Armstrong, Phys.\ Rev. {\bo D62}, 042002 (2000).

\n
\hangindent=0.2in
\hangafter=1
[4] A. Vecchio \& C. Cutler, in {\it Laser Interferometer Space Antenna: Second INternational LISA Symposium on the Detection and Obesrvation of Gravitational Waves in Space}  AIP Conference Proceedings 456, 101 (1998).

\n
\hangindent=0.2in
\hangafter=1
[5]  M. Tinto, J.W. Armstrong, \& F.B. Estabrook, Phys.\ Rev. {\bo D}, accepted (2000).

\n
\hangindent=0.2in
\hangafter=1
[6]  F.D. Cady \& R.W. Hellings, manuscript in preparation (2001).

\n
\hangindent=0.2in
\hangafter=1
[7]  R.W. Hellings, {\it et.\ al}, Optics Comm. {{\bo 124} 313-320 (1996).

\bye